\documentclass[12pt,a4paper]{article}
\usepackage[utf8]{inputenc}
\usepackage{amsmath}
\usepackage{amsfonts}
\usepackage{amssymb}
\usepackage{makeidx}
\usepackage{graphicx}

\begin{document}

\begin{center}
{\Large \bf Tunneling time from locally periodic potential in space fractional quantum mechanics
  }

\vspace{1.3cm}

{\sf   Mohammad Hasan  \footnote{e-mail address: \ \ mhasan@isro.gov.in, \ \ mohammadhasan786@gmail.com}$^{,3}$,
 Bhabani Prasad Mandal \footnote{e-mail address:
\ \ bhabani.mandal@gmail.com, \ \ bhabani@bhu.ac.in  }}

\bigskip

{\em $^{1}$Indian Space Research Organisation,
Bangalore-560094, INDIA \\
$^{2,3}$Department of Physics,
Banaras Hindu University,
Varanasi-221005, INDIA. \\ }

\bigskip
\bigskip

\noindent {\bf Abstract}

\end{center}
We calculate the time taken by a wave packet to travel through a classically forbidden locally periodic rectangular potential in space fractional quantum mechanics (SFQM). We obtain the  close form expression of tunneling time from such a potential by stationary phase  method. We show that tunneling time depends upon the width $b$ of the single barrier and separation $L$ between the barriers in the limit  $b \to \infty$ and therefore generalized Hartman effect doesn't exist in SFQM. We observe that in SFQM,  the tunneling time for large $b$ in the case of locally periodic potential is smaller than the tunneling from a single barrier of the same width $b$. It is further shown that with the increase in barrier numbers, the tunneling time reduces in SFQM in the limit of large $b$.

\medskip
\vspace{1in}
\newpage

\section{Introduction}
In quantum mechanics the concept of fractals were first introduced in its path integral (PI) formulation \cite{feynman}. In the PI formulation of quantum mechanics, the path integrals are taken over Brownian-like quantum mechanical paths that results in Schrodinger equation. However the Brownian process (the random processes governed by Gaussian probability distribution) are the subset of a more general class of process called Levy $\alpha$-stable random process. The Levy $\alpha$-stable random process are non-Gaussian process often characterized by Levy index $\alpha$, $0 < \alpha \leq 2$. For $\alpha=2$, the Levy process are transformed to Brownian process or in other words the Levy paths are Brownian paths for $\alpha=2$. The path integral approach of quantum mechanics was generalized by Nick Laskin by considering the path integrals over Levy flight paths \cite{1,2}. The PI formulation over Levy paths leads to fractional Schrodinger equation \cite{3} and the corresponding branch of quantum mechanics is known as space fractional quantum mechanics (SFQM). Owing to its natural generalization of standard quantum mechanics, the field of SFQM has grown fast over the last one and a half decade.  Various applications of fractional generalization of quantum mechanics have also been discussed \cite{4,5,6} including its applications in fractional quantum optics \cite{longhi_fractional_optics} and diffraction-free beams in SFQM \cite{diffraction_free_beam_nature}. New scattering features from non-Hermitian potential in SFQM have also been reported \cite{nh_sfqm}.
\paragraph{}        
The particle tunneling from a classically forbidden region is one of the most of fundamental problem of quantum mechanics. It is also one of the few earliest studied problems of quantum mechanics  started in the year $1928$ \cite{ nordheim1928, gurney1928}. Since then the quantum mechanical tunneling has long been studied \cite{condon, wigner_1955, david_bohm_1951}. In the year $1962$, Hartman studied the time taken by a wave-packet to tunnel through the classically forbidden region imposed by metal-insulator-metal sandwich \cite{hartman_paper} by stationary phase method. It was found that the tunneling time doesn't depend upon the thickness of tunneling region. This famous paradox is known as Hartman effect. Later independent study by Fletcher showed the same effect i.e. the tunneling time of evanescent wave shows saturation with the barrier thickness \cite{fletcher}. In later years several studies were done to understand the nature of tunneling time. The numerical monitoring of time evolution of the tunneling particle have shown that the tunneling time agrees with the ones obtained by stationary phase method \cite{aquino_1998}. It was also found from stationary phase method that in case of two opaque barrier separated by a finite distance, the tunneling time is also independent of the separation in the limit when the thickness of the barrier is large. This phenomena was termed as generalized Hartman effect \cite{generalized_hartman}. For multi-barrier tunneling, the tunneling phase time is also independent of barrier thickness and inter-barrier separation \cite{esposito_multi_barrier}. The close form expressions of super-lattice tunneling time have been obtained in \cite{pereyra_2000} and is shown that the super-lattice tunneling time can be smaller than the free motion time when particle energy lies in the energy gap of the super-lattice structure. The problem of tunneling time was also studied for complex potentials. For complex potentials associated with elastic and inelastic channels, the tunneling time was found to saturate with the thickness of the barrier for the case of weak absorptions \cite{dutta, our}.     
\paragraph{}
To test the validity of the theoretical results of tunneling time, a series of accurate measurements were conducted by different researchers. The measurement of single photon tunneling time have revealed the superluminal nature of the tunneling phenomena \cite{sl_prl}.  The experiments were done with microwave \cite {nimtz,ph,ragni} as well as in optical ranges \cite {sattari,longhi1,olindo}. The tunneling time in all such experiments doesn't found to be changing with the thickness of the tunneling region. The tunneling time was also studied with double barrier optical gratings \cite{longhi1} and double barrier photonic band gaps \cite{longhi2}. The observed tunneling time was paradoxically short. The results of these experiments also favored the generalized Hartman effect as the tunneling time was found to be independent of the gap between the two barrier distance. For discussions on the validity of the generalized Hartman effect, the readers are referred to some of the critical articles/comments : \cite{questions_ghf1,questions_ghf2,questions_ghf3}.
\paragraph{}
Considering the growing interest and development of SFQM as well as in the tunneling time, we explore the tunneling time by stationary phase method in the domain of SFQM for locally periodic rectangular potential. The tunneling time from an opaque rectangular barrier in SFQM have been discussed in \cite{tt_sfqm} and is shown that the Hartman effect doesn't exist in SFQM. In this paper we provide close form expression of tunneling time from a locally periodic potential in SFQM. We show that the generalized Hartman effect doesn't exist in SFQM. However this effect is restored as a special case for standard quantum mechanics ($\alpha =2$). We found that the tunneling time in SFQM for locally periodic potential (of arbitrary periodicity $N$) each of width $b$ and separated by a distance $L$ is smaller as compared to single barrier of width $b$ in the limit  $b \rightarrow \infty$. We also show that in the limit $b \rightarrow \infty$, the tunneling time reduces as the number of barriers increases. The tunneling time first attain a maxima with increasing barrier thickness and then begin to decrease as the barrier thickness further increases. The decrease in tunneling time for large barrier have also been shown  to occur in standard quantum mechanics with wave packet of large momentum spreads \cite{tt_decrease}.
\paragraph{}
We organized the paper as follows: in  section \ref{intro_tt} we outline  the stationary phase  method of calculating the tunneling time and the Hartman effect of standard quantum mechanics. We discuss the fractional Schrodinger equation in \ref{fse}. In section \ref{tt_in_sfqm} we calculate the tunneling time from classically opaque locally periodic rectangular potential. We also discuss here (in sub-section \ref{2_barrier}) the special case of double barrier potential in SFQM and show that generalized Hartman effect doesn't exist in the domain of SFQM. The limiting case of tunneling time for locally periodic potential for large $b$ is also discussed separately in sub-section \ref{n_barrer}. Finally the results are discussed in section \ref{results_discussions}.

\section{Tunneling time in standard quantum mechanics}
\label{intro_tt}
In this section we discuss the calculation of tunneling time by stationary phase method  for a particle traversing a classically forbidden region of width $b$ \cite{dutta_roy_book}. According to stationary phase method, the time difference between the peak of the incoming and outgoing wave packet gives the tunneling time as the wave packet crosses the potential barrier. We briefly present the methodology of stationary phase method below. 
\paragraph{}
Consider a localized wave packet $W_{k_{0}} (k)$ which is a normalized Gaussian function having peak at the mean momentum $ \hbar k_{0}$. Here wave number $k=\sqrt{2mE}$. The time evolution of $W_{k_{0}} (k)$ is given by 
\begin{equation}
\int W_{k_{0}} (k)e^{i(kx-\frac{Et}{\hbar})}dk
\label{localized_wave_packet}
\end{equation}
This wave packet propagates towards positive $x$ direction. After the transmission from the potential barrier the emerged wave packet would be
\begin{equation}
\int W_{k_{0}} (k) \vert t(k) \vert e^{i(kx-\frac{Et}{\hbar} +\Phi(k))}dk
\label{emerged_wave_packet}
\end{equation}
Here $t(k)$ is the transmission coefficient through the potential barrier $V(x)$ ($V(x)=V $ for $0 \leq x \leq b$ and zero elsewhere). According to SPM, the tunneling time $\tau$ is
\begin{equation}
\frac{d}{dk} \left( kb-\frac{E\tau}{\hbar} +\Phi(k) \right)=0
\label{spm_condition}
\end{equation}
The above condition provides the expression for tunneling time as 
\begin{equation}
\tau= \hbar \frac{d \Phi(E)}{dE} +\frac{b}{(\frac{\hbar k}{m})}
\label{phase_delay_time}
\end{equation}
For a square barrier of width $b$, the expression for tunneling time is \cite{dutta_roy_book}
\begin{equation}
\tau= \hbar \frac{d}{dE} \tan^{-1} \left( \frac{k^{2}-q^{2}}{2kq} \tanh{qb}\right)
\end{equation}
Here $q= \sqrt{2m(V-E)}/\hbar$. We observe $\tau \rightarrow 0$ as $b \rightarrow 0$. This is expected. However when $b \rightarrow \infty$ we observed $\tau \rightarrow \frac{2m}{\hbar qk}$. This expression doesn't involve $b$ i.e. tunneling time is independent of the width of the barrier  for a sufficiently opaque barrier. This is known as the famous Hartman effect, i.e for a sufficiently thick barrier, the tunneling time is independent of the width of the barrier. In the system of units  $2m=1$, $\hbar=1$, $c=1$ the expression of tunneling time for $b \rightarrow 0$ becomes
\begin{equation}
\lim_{b\rightarrow \infty} \tau = \frac{1}{qk}
\label{tt_qm}
\end{equation}

\section{The fractional Schrodinger equation}
\label{fse}
The fractional Schrodinger equation in one dimension is 
\begin{equation}
i \hbar \frac{\partial \psi (x,t)}{\partial t}= H_{\alpha} (x,t) \psi(x,t) 
\label{tdfse}
\end{equation}
where $1 < \alpha \leq 2$ (see discussion in \cite{1} for range of $\alpha$). The fractional Hamiltonian operator $H_{\alpha} (x,t)$ is 
\begin{equation}
H_{\alpha} (x,t)=D_{\alpha} (-\hbar^{2} \Delta)^{\frac{\alpha}{2}} +V(x,t)
\end{equation}
In the above $D_{\alpha}$ is a constant. It depends on the system characteristics. $\Delta=\frac{\partial^{2}}{\partial x^{2}}$ . The Riesz fractional derivative $(-\hbar^{2} \Delta)^{\alpha/2}$ of a function $\Psi(x,t)$ is defined through its Fourier transform $\tilde{\Psi}(p,t)$ as
\begin{equation}
(-\hbar^{2}\Delta)^{\frac{\alpha}{2}} \Psi(x,t)=\frac{1}{2\pi \hbar} \int_{-\infty} ^{\infty} { \tilde{\Psi}(p,t)\vert p \vert ^{\alpha} e^{\frac{ipx}{\hbar}}dp } 
\label{Riesz_fractional_derivative}
\end{equation}
If the potential $V(x,t)$ is independent of time the fractional Hamiltonian operator $H_{\alpha}(x)=D_{\alpha} (-\hbar^{2} \Delta)^{\frac{\alpha}{2}} +V(x)$ is time independent as well. In this case we have the time independent fractional Schrodinger equation as
\begin{equation}
D_{\alpha} (-\hbar^{2} \Delta)^{\frac{\alpha}{2}}\psi(x)+V(x)\psi(x)=E\psi(x)
\label{tifse}
\end{equation} 
where $E$ is particle's energy and $\psi(x,t)=\psi(x)e^{-\frac{iEt}{\hbar}}$. For a detail discussion on fractional Schrodinger equation, readers are referred to \cite{3}.
\section{Tunneling time from locally periodic potential in space fractional quantum mechanics}
\label{tt_in_sfqm}
To calculate the tunneling time for a potential, we have to find the phase of the transmission coefficient. This phase is then differentiated with respect to energy in order to obtain the phase delay time. This time combined with the free passage time of the spatial extent of the potential gives the net tunneling time from the given potential distribution.  As the tunneling from a periodic system depend upon the characteristics of the unit cell potential, we prefer to start our discussion of finding the phase of the transmission coefficient of the periodic system in terms of the parameters of the unit cell rectangular barrier. This also help to compare the tunneling time from locally periodic system as compared to the single barrier case in certain limit (such as thickness of unit cell $b \rightarrow \infty $). The detail of the calculations  are illustrated below along with the separations of real and imaginary parts of various quantities of interests. \\

The transmission coefficient for a square barrier potential $V(x)=V$ confined over the region $0 \leq x \leq b$ and zero elsewhere has been calculated in Ref \cite{4}. Ref. \cite{tare} provide transfer matrix for a such a barrier. For a single barrier of width $b$, the transmission coefficient is
\begin{equation}
t_{1}=\frac{1}{M_{1}}
\end{equation}
where,
\begin{equation}
M_{1}=(\cos{\overline{k}_{\alpha}b}- i\mu \sin{\overline{k}_{\alpha}b}) e^{ik_{\alpha}b}
\label{m1}
\end{equation}
In the above expression,
\begin{equation}
k_{\alpha}=\left( \frac{E}{D_{\alpha} \hbar^{\alpha}}\right)^{\frac{1}{\alpha}}
\end{equation}
and
\begin{equation}
\overline{k}_{\alpha}=\left( \frac{E-V}{D_{\alpha} \hbar^{\alpha}}\right)^{\frac{1}{\alpha}}
\end{equation}
Eq. \ref{m1} can be written as
\begin{equation}
M_{1}=\sqrt{v_{\alpha}} e^{-i\delta}
\end{equation}
where $\delta=\theta -k_{\alpha} b$. For classically forbidden case $E<V$, the expressions for $v_{\alpha}$ and $\theta$ are  
\begin{multline}
v_{\alpha}=\frac{1}{16}\Big[ \{ 8-\varepsilon_{-}^{2}-\varepsilon_{+}^{2} -(\varepsilon_{+}^{2}- \varepsilon_{-}^{2})\cos{2\beta} \}\cos{2\eta} + \\
\{ 8+\varepsilon_{-}^{2}+\varepsilon_{+}^{2} +(\varepsilon_{+}^{2}- \varepsilon_{-}^{2})\cos{2\beta}\} \cosh{2\xi}   \\
-8\varepsilon_{-}\sin{\beta}\sin{2\eta}+ 8\varepsilon_{+}\cos{\beta}\sinh{2\xi} \Big]
\label{valpha}
\end{multline}
and,
\begin{equation}
 {\theta}= \tan^{-1} \Big (  \frac{ 2\varepsilon_{\alpha} \sin{\eta}\sinh{\xi} +(\varepsilon_{\alpha}^{2}+ 1) \sin{\eta}\cosh{\xi} \cos{\beta }  +(\varepsilon_{\alpha}^{2}- 1) \cos{\eta}\sinh{\xi} \sin{\beta }
 } {2\varepsilon_{\alpha} \cos{\eta}\cosh{\xi} +(\varepsilon_{\alpha}^{2}+ 1) \cos{\eta}\sinh{\xi} \cos{\beta } -(\varepsilon_{\alpha}^{2}- 1) \sin{\eta}\cosh{\xi} \sin{\beta } } \Big )
\end{equation}
In theses expressions $q_{\alpha}$, $\varepsilon_{\alpha}$ and $\varepsilon_{\pm}$ are defined as
\begin{equation}
q_{\alpha}=\left( \frac{V-E}{D_{\alpha}}\right)^{\frac{1}{\alpha}}
\label{q_alpha}
\end{equation}
\begin{equation}
\varepsilon_{\alpha}=\left(  \frac{k_{\alpha}}{q_{\alpha}} \right)^{\alpha-1}
\label{eps_alpha}
\end{equation}
\begin{equation}
\varepsilon_{\pm}=\varepsilon_{\alpha}\pm\frac{1}{\varepsilon_{\alpha}}
\label{eps_plus_minus}
\end{equation}
and,
\begin{equation}
\eta=q_{\alpha}b\cos{\frac{\pi}{\alpha}} , \quad   \xi=q_{\alpha}b\sin{\frac{\pi}{\alpha}} , \quad   \beta=\frac{\alpha-1}{\alpha}\pi, \quad  \gamma=\frac{\pi}{\alpha}
\end{equation} 
The transmission coefficient from a locally periodic rectangular potential in SFQM is \cite{tare}
\begin{equation}
t_{N}=\frac{e^{-ik_{\alpha} N s}}{M_{N}}
\end{equation}
where $N$ is the number of barriers and 
\begin{equation}
M_{N}= M_{1} e^{-ik_{\alpha}s} U_{N-1} (\chi)-U_{N-2} (\chi)
\end{equation}
Here $s=b+L$, $L$ being the separation between two consecutive rectangular barriers. And,
\begin{equation}
\chi=\sqrt{v_{\alpha}} \cos{(\delta +k_{\alpha}s)}
\end{equation}
To find the phase of $t_{N}$, we separate $M_{N}$ in real and imaginary parts,
\begin{equation}
M_{N}=P_{N}-iQ_{N}
\end{equation}
where,
\begin{equation}
P_{N}=\chi U_{N-1} (\chi)-U_{N-2} (\chi)
\end{equation}
\begin{equation}
Q_{N}= \sqrt{v_{\alpha} -\chi^{2}} U_{N-1} (\chi) 
\end{equation}
Thus the phase of $t_{N}$ is found to be
\begin{equation}
\zeta =\Phi - k_{\alpha} Ns
\end{equation}
where,
\begin{equation}
\Phi= \tan^{-1}{\left (\frac{Q_{N}}{P_{N}}\right )}
\end{equation}

Now we have to evaluate $\frac{d\Phi }{dE}$ to find the phase delay time. After performing lengthy algebra, it can be shown that
\begin{equation}
\frac{d\Phi}{dE}=\frac{A_{1}}{A_{2}}
\label{dPhi_dE}
\end{equation}
Where $A_{2}$ is,
\begin{equation}
A_{2}= v_{\alpha} U_{N-1}^{2}(\chi) + U_{N-2}^{2}(\chi)-2 \chi U_{N-1}(\chi) U_{N-2}(\chi)
\label{a2}
\end{equation}
The expression for  $A_{1}$ is mathematically lengthy and we define it, keeping the future calculation in mind, through the use of following variables:
\begin{equation}
A_{1}= a_{1}+a_{2}
\label{a1}
\end{equation}
with $a_{1}$ and $a_{2}$ defined below.
\begin{equation}
a_{1}= \frac{1}{2 \sqrt{v_{\alpha}-\chi^{2}}} \left( B_{1} -2 B_{2}\right)
\label{small_a1}
\end{equation}
\begin{equation}
a_{2}= \frac{\sqrt{v_{\alpha}-\chi^{2}}}{\chi ^{2}-1} \left( B_{3}-B_{4}\right)
\label{small_a2}
\end{equation}
Various $B_{i}s$  are given below
\begin{equation}
B_{1}=U_{N-1}(\chi)( v_{\alpha}'-2\chi \chi') T_{N}(\chi)
\end{equation}
\begin{equation}
B_{2}= (v_{\alpha} -\chi ^{2}) \chi' U_{N-1}^{2}(\chi)
\end{equation}
\begin{equation}
B_{3}= \chi 'U_{N-2}(\chi)(N U_{N-2}(\chi)-\chi U_{N-1}(\chi) )
\end{equation}
\begin{equation}
B_{4}= (N-1) \chi'U_{N-1}(\chi) U_{N-3}(\chi)
\end{equation}
In the above,
\begin{equation}
\chi'=\frac{v_{\alpha}'}{2 \sqrt{v_{\alpha}}} \cos{(\delta +k_{\alpha}s)} -\sqrt{v_{\alpha}} (\delta ' +k_{\alpha}'s) \sin{(\delta +k_{\alpha}s)} 
\label{chi_prime}
\end{equation}
Further in Eq. \ref{chi_prime},
\begin{equation}
k_{\alpha}'=\frac{dk_{\alpha}}{dE}=\frac{k_{\alpha}^{1-\alpha}}{\alpha D_{\alpha}}
\label{dkalpha_dk}
\end{equation}
and,
\begin{equation}
\delta '=\frac{d \delta}{dE}
\end{equation}
The expression for $\delta '$ can be evaluated as
\begin{equation}
\delta'=\left(\frac{d_{\alpha}}{v_{\alpha}}-\frac{b k_{\alpha}^{1-\alpha}}{\alpha D_{\alpha}} \right)
\end{equation}
where $v_{\alpha}$ is given by Eq. \ref{valpha} and $d_{\alpha}$ is defined below
\begin{multline}
d_{\alpha}= \frac{1}{2}b \varepsilon_{+}q'_{\alpha}\cos{\beta}\cos{\gamma}\cosh{2\xi}+\frac{1}{2}b \varepsilon_{-}q'_{\alpha}\sin{\beta}\sin{\gamma}\cos{2\eta}\\ 
+\frac{1}{4}\varepsilon'_{+}\cos{\beta}\sin{2\eta}+\frac{1}{2}b q'_{\alpha}(\cos{\gamma}\sinh{2\xi}+\sin{\gamma}\sin{2\eta})\\
+\frac{1}{8}b q'_{\alpha}(\sinh{2\xi}\cos{\gamma}-\sin{2\eta}\sin{\gamma})(\varepsilon_{+}^{2}\cos^{2}{\beta}+ \varepsilon_{-}^{2}\sin^{2}{\beta})\\
+\frac{1}{8}\sin{2\beta}(\cosh^{2}{\xi}\sin^{2}{\eta}+\sinh^{2}{\xi}\cos^{2}{\eta})(\varepsilon_{+}\varepsilon'_{-}-\varepsilon_{-}\varepsilon'_{+})\\
+\frac{1}{4}\varepsilon'_{-}\sin{\beta}\sinh{2\xi}
\label{d_alpha}
\end{multline} 
and,
\begin{equation}
v_{\alpha}'=\frac{d v_{\alpha}}{dE}=v_{1\alpha}+bq_{\alpha}' v_{2\alpha}
\label{dvalpha_dk}
\end{equation}
where,

\begin{multline}
v_{1\alpha}= \frac{1}{4}(\cosh{2\xi}-\cos{2\eta}) (\varepsilon_{+} \varepsilon_{+}' \cos^{2}{\beta}+ \varepsilon_{-} \varepsilon_{-}' \sin^{2}{\beta}) \\ -\frac{1}{2} \varepsilon_{-}' \sin{\beta} \sin{2 \eta} +\frac{1}{2} \varepsilon_{+}' \cos{\beta} \sinh{2 \xi} 
\end{multline}
\begin{multline}
v_{2\alpha}= \frac{1}{4} (\sin{2 \eta} \cos{\gamma}+ \sinh{2 \xi} \sin{\gamma}) ( \varepsilon_{+}^{2} \cos ^{2}{\beta} +\varepsilon_{-}^{2} \sin ^{2}{\beta}) + \\ (\varepsilon_{+} \cos{\beta} \cosh{2 \xi} \sin{\gamma} -\varepsilon_{-} \sin{\beta} \cos{2 \eta} \cos{\gamma}) +\\
(\sin{\gamma} \sinh{2 \xi}-\cos{\gamma} \sin{2 \eta})
 \end{multline}

In the above equations
\begin{equation}
q'_{\alpha}=\frac{dq_{\alpha}}{dE}=-\frac{1}{\alpha D_{\alpha}}q_{\alpha}^{1-\alpha}
\label{q_dash}
\end{equation}

\begin{equation}
\varepsilon'_{\pm}=\frac{d \varepsilon_{\pm} }{dE}=\frac{\alpha-1}{\alpha} \frac{V}{(V-E)^{2}} \varepsilon_{\alpha}^{\frac{1}{1-\alpha}} (1\mp\varepsilon_{\alpha}^{-2})
\label{eps_plus_dash}
\end{equation}
Now the tunneling time $\Gamma _{\alpha}^{N}$ from locally periodic rectangular barrier potential can be obtained as
\begin{equation}
\Gamma_{\alpha}^{N}=\frac{d \Phi}{dE}- N s k_{\alpha}'+ \frac{(N-1)s+b}{2k}
\label{tt_general_n_sfqm}
\end{equation}
where $\frac{d \Phi }{dE}$ is given by Eq. \ref{dPhi_dE}.
\subsection{Special case: double barrier potential}
\label{2_barrier}
To investigate the generalized Hartman effect, we take the case of $n=2$. In this case we have
\begin{equation}
P_{2}=2 \chi ^{2}-1
\end{equation}
\begin{equation}
Q_{2}=2 \chi \sqrt{v_{\alpha}-\chi^{2}}
\end{equation}
On performing the necessary algebra, the tunneling time from double barrier potential is obtained as
\begin{equation}
\Gamma_{\alpha}^{2}= Z_{\alpha}- \frac{2s k_{\alpha}^{1-\alpha}}{\alpha D_{\alpha}} +\frac{s+b}{2k}
\end{equation}
where,
\begin{equation}
Z_{\alpha}=\frac{v_{\alpha}' \chi (2 \chi^{2}-1)-2\chi'(2v_{\alpha} \chi^{2}-2 \chi^{2}+v_{\alpha})}{\sqrt{v_{\alpha}-\chi^{2}} (4 \chi^{2}v_{\alpha}-4 \chi^{2}+1)}
\label{z_alpha}
\end{equation}
We take the limiting case
\begin{equation}
\lim_{b \to \infty} v_{\alpha} \sim e^{2\xi} f_{1}
\label{valpha_inf}
\end{equation}
\begin{equation}
\lim_{b \to \infty} v_{\alpha}' \sim e^{2\xi} (f_{2}+bf_{3})
\label{vdash_alpha_inf}
\end{equation}
where,
\begin{equation}
f_{1}= 4\varepsilon_{+}\cos{\beta} + \frac{1}{32} \{ 8 +\varepsilon_{+}^{2} +\varepsilon_{-}^{2} +(\varepsilon_{+}^{2} -\varepsilon_{-}^{2})\cos{2\beta} \}
\label{f1} 
\end{equation}
\begin{equation}
f_{2}= \frac{1}{8}(2 \varepsilon_{+}'\cos{\beta}+\varepsilon_{-}\varepsilon_{-}'\sin^{2}{\beta}+\varepsilon_{+}\varepsilon_{+}'\cos^{2}{\beta})
\label{f2} 
\end{equation}
\begin{equation}
f_{3}=\frac{1}{8} q_{\alpha}'\sin{\gamma} \left[ 4 \varepsilon_{+}\cos{\beta}+ \varepsilon_{+}^{2} \cos^{2}{\beta}+\varepsilon_{-}^{2} \sin^{2}{\beta} +4 \right] 
\label{f3} 
\end{equation}
In terms of $f_{1}$, $f_{2}$ and $f_{3}$ we evaluate,
\begin{equation}
\lim_{b \to \infty} \chi' \sim \frac{e^{\xi}}{2 \sqrt{f_{1}}} (f_{2}+b f_{3}) \cos{(\delta + k_{\alpha} s)} - e^{\xi} \sqrt{f_{1}} (\delta' + k_{\alpha}' s) \sin{(\delta + k_{\alpha} s)}
\label{chidash_alpha_inf}
\end{equation}
Considering the dominant term of Eq. \ref{z_alpha} the limiting case $b \rightarrow \infty$ is simplified to 
\begin{equation}
\lim_{b \to \infty} Z_{\alpha} \sim \delta'+k_{\alpha}'s
\label{chidash_alpha_inf}
\end{equation}
Hence for double barrier case
\begin{equation}
\lim_{b \to \infty} \Gamma_{\alpha}^{2} \sim \delta'+k_{\alpha}'s - \frac{2s k_{\alpha}^{1-\alpha}}{\alpha D_{\alpha}} +\frac{s+b}{2k} 
\end{equation}
We recognise that $\delta'$ is the phase delay time from a single rectangular barrier of width $b$ in SFQM. Thus the term $\delta' +\frac{b}{2k}=\tau_{\alpha}$ is the net tunneling time from single barrier potential in SFQM. Hence we arrive at
\begin{equation}
\lim_{b \to \infty} \Gamma_{\alpha}^{2} \sim  \left(\lim_{b \to \infty} \tau_{\alpha} \right)+ s\left( \frac{1}{2k} - \frac{1 }{\alpha D_{\alpha} k_{\alpha}^{\alpha-1}} \right) 
\label{gamma_alpha_large_b}
\end{equation}
We see from Eq. \ref{gamma_alpha_large_b} that the tunneling time does depend upon $s=b+L$ for $\alpha \neq 2$. Hence the generalized Hartman effect doesn't exist in SFQM. This is further observed that the second term in parenthesis of Eq. \ref{gamma_alpha_large_b} 
\begin{equation}
w_{\alpha}= \frac{1}{2k} - \frac{1 }{\alpha D_{\alpha} k_{\alpha}^{\alpha-1}} 
\label{zalpha}
\end{equation}
vanishes for $\alpha=2$ and we restore the generalized Hartman effect of standard quantum mechanics (as $\lim_{b \to \infty} \tau_{\alpha=2}=\frac{1}{qk}$). Further we see that $w_{\alpha} \leq 0$ for $1 < \alpha  \leq 2$. Therefore
\begin{equation}
\lim_{b \to \infty} \Gamma_{\alpha}^{2} <  \lim_{b \to \infty} \tau_{\alpha} 
\label{gamma_alpha_less_than_tau_alpha}
\end{equation}
for $1 <\alpha < 2$. This shows that in SFQM, the tunneling time for double barrier potential is smaller than the tunneling time for single barrier potential of large width $b$.  
\subsection{$b \rightarrow \infty$ case of locally periodic potential}
\label{n_barrer}
\begin{figure}
\begin{center}
\includegraphics[scale=0.35]{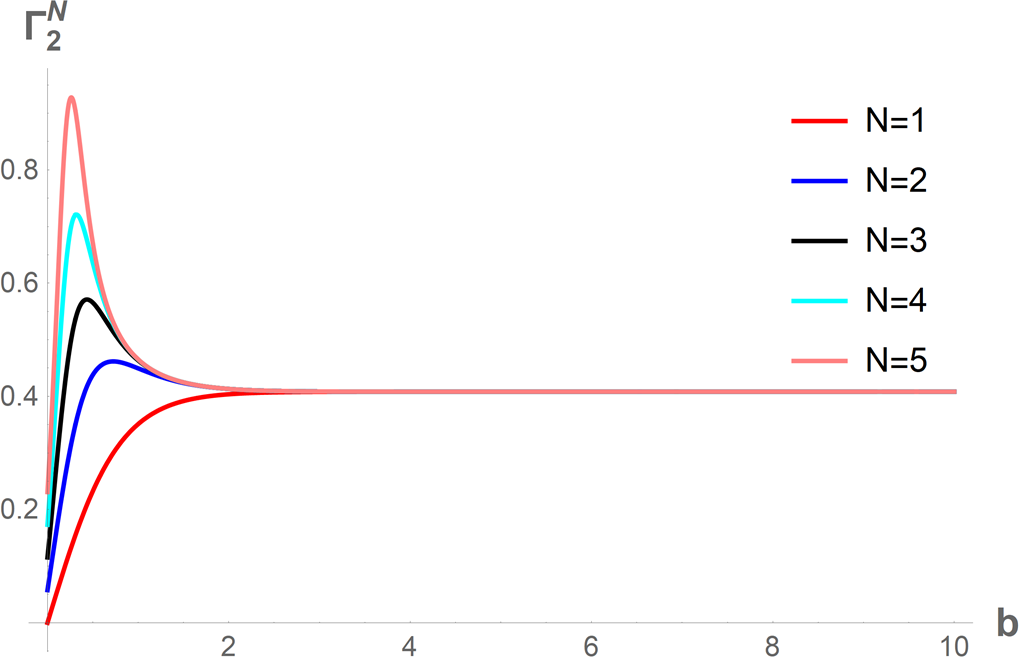} a  \\ 
\includegraphics[scale=0.35]{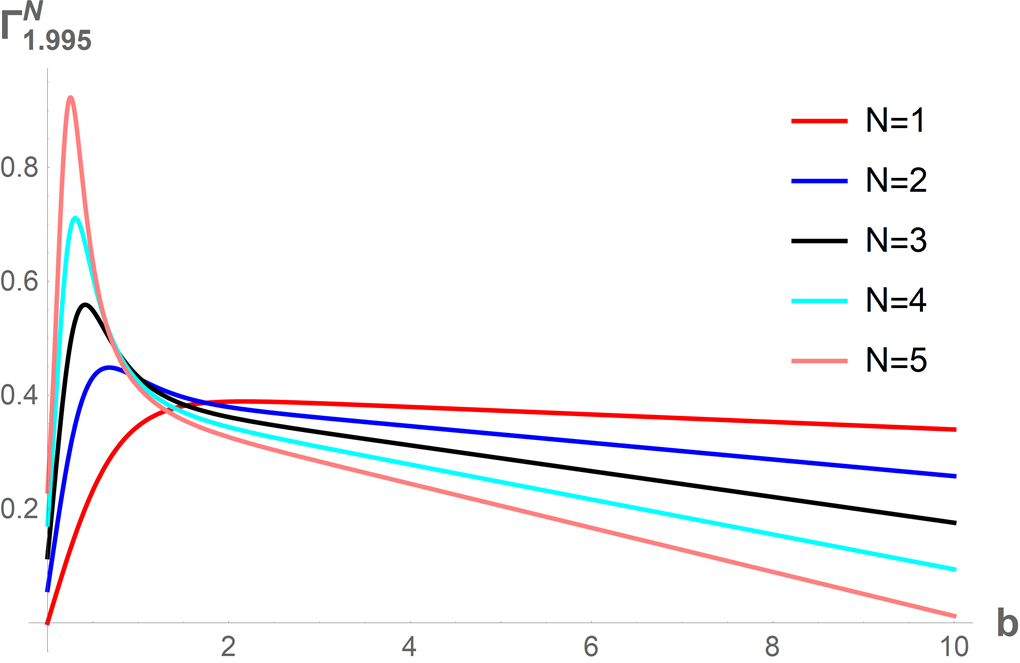} b
\caption{\it Plots showing the variation of tunneling time with `$b$' for different number of barriers for $\alpha =2$ (Fig-a). The Hartman effect and generalized Hartman effect is clearly seen here. Fig -b shows the variation of tunneling time with `$b$' for $\alpha =1.995$ for different number of barriers. It is seen that for large $b$, the tunneling time for multibarrier case is smaller than the case of single barrier. Further as the number of barrier increases, the tunneling time reduces for large $b$ as compared to the case of smaller number of barriers. All these results are consistent with the theoretical formulation developed in the paper. For the above figures, potential height $V=5$,energy $E=3$, separation between each barrier $L=0.2$ and $v=10^{-4}$.  }  
\label{tt_graph_n_barrier}
\end{center}
\end{figure}

Next we study the case of large width $b$ for a general $N$. Due to the presence of several Chebyshev polynomials  which are oscillatory in nature, the computation of $b \rightarrow \infty$ is somewhat tricky. Therefore we illustrate the various steps of calculations below.
We note that for large $b$, the dominant term of $A_{1}$ and $A_{2}$ is $U_{N-1} (\chi) ^{2}$. Therefore we divide $A_{1}$ and $A_{2}$ by $U_{N-1} (\chi)^{2}$ and evaluate the limiting case $b \rightarrow \infty$.  We proceed as follows:
\begin{equation}
\lim_{b \to \infty} \left( \frac{a_{1}}{U_{N-1} (\chi)^{2}} \right)= \lim_{b \to \infty} \left( \frac{( v_{\alpha}'-2\chi \chi') }{2 \sqrt{v_{\alpha}-\chi^{2}}} \right) \times \left( \lim_{b \to \infty} \frac{T_{N}(\chi)}{U_{N-1}(\chi)} \right) - \lim_{b \to \infty}  (\sqrt{v_{\alpha}-\chi^{2}}) \chi '
\end{equation}
The limiting case of the fraction containing the Chebyshev polynomials goes as $\chi$ as $ b \rightarrow \infty$. Further simplification leads to
\begin{equation}
\lim_{b \to \infty} \left( \frac{a_{1}}{U_{N-1} (\chi)^{2}} \right) \sim \lim_{b \to \infty} \left( \frac{ v_{\alpha}' \chi -2 v_{\alpha} \chi '}{2 \sqrt{v_{\alpha}-\chi^{2}}} \right)
\label{b_large_small_a1}
\end{equation}
Similarly we can show that 
\begin{equation}
\lim_{b \to \infty} \left( \frac{a_{2}}{U_{N-1} (\chi)^{2}} \right)= 0
\label{b_large_small_a2}
\end{equation}
and,
\begin{equation}
\lim_{b \to \infty} \left( \frac{A_{2}}{U_{N-1} (\chi)^{2}} \right)\sim v_{\alpha}
\label{b_large_a2}
\end{equation}
Thus,
\begin{equation}
\lim_{b \to \infty} \left( \frac{A_{1}}{A_{2}} \right) \sim \lim_{b \to \infty} \left( \frac{ v_{\alpha}' \chi -2 v_{\alpha} \chi '}{2 v_{\alpha} \sqrt{v_{\alpha}-\chi^{2}}} \right)
\end{equation}
Writing $\sqrt{v_{\alpha}-\chi^{2}}=\sqrt{v_{\alpha}} \sin{(\delta+ k_{\alpha}s)}$ and using Eq. \ref{chi_prime} (expression for $\chi '$) it can be shown that
\begin{equation}
\lim_{b \to \infty} \left( \frac{A_{1}}{A_{2}} \right) \sim \lim_{b \to \infty} (\delta'+ k_{\alpha}' s)
\label{b_large_a1_by_a2}
\end{equation}
Using the above result in Eq. \ref{tt_general_n_sfqm} and recognising $\tau_{\alpha}= \delta' +\frac{b}{2k}$ we get the limiting behaviour of the tunneling time as
\begin{equation}
\lim_{b \to \infty} \Gamma_{\alpha}^{N} \sim  \left(\lim_{b \to \infty} \tau_{\alpha} \right)+ (N-1)s w_{\alpha}
\label{gamma_alpha_n_large_b}
\end{equation}
The above expression correctly reduces to Eq. \ref{gamma_alpha_large_b} for double barrier case ($N=2$). Also for $\alpha=2$, $w_{\alpha}=0$ and we observe

\begin{equation}
\lim_{b \to \infty} \Gamma_{\alpha =2}^{N} =  \lim_{b \to \infty} \tau_{\alpha =2} 
\end{equation}
which is again generalized Hartman effect of standard quantum mechanics. Further as $w_{\alpha} <0$ for $1 < \alpha <2$, hence
\begin{equation}
\lim_{b \to \infty} \Gamma_{\alpha}^{N} <  \lim_{b \to \infty} \tau_{\alpha} 
\end{equation}
This shows that in SFQM, the tunneling time from a locally periodic potential with large width of the unit cell is smaller than the tunneling time from the single unit cell. From Eq. \ref{gamma_alpha_n_large_b} it is seen that for a pair of natural numbers $N_{1}$ and $N_{2}$ such that $N_{1} < N_{2}$,
\begin{equation}
\lim_{b \to \infty} \Gamma_{\alpha}^{N_{2}} <  \lim_{b \to \infty} \Gamma_{\alpha}^{N_{1}} 
\end{equation}
for $1 < \alpha <2$. Therefore in the limit $b \rightarrow \infty$, the tunneling time reduces as the number of barrier increases (in SFQM). This is also graphically shown in Fig \ref{tt_graph_n_barrier}-b.

\section{Results and Discussions}
\label{results_discussions}
We have provided the close form expression of tunneling time from locally periodic rectangular potential in space fractional quantum mechanics (SFQM). In the limiting of case of large width $b$ of the unit cell barrier, we see that the tunneling time depends upon the width $b$ for $1< \alpha < 2$ for arbitrary $N$ number of barriers. Therefore the generalized Hartman effect doesn't exist in SFQM. In general, the tunneling time depends upon spatial extent of the potential for any extent (small or large) in SFQM.  The generalized Hartman effect is restored for $\alpha=2$ case i.e. the standard quantum mechanics. 
\paragraph{}
It is observed that, in SFQM, the tunneling time from a locally  periodic rectangular potential made of `large width' of the unit cell potential is smaller than the tunneling time of the same unit cell potential. Furthermore we have shown that in SFQM, in general, in the limit $b \rightarrow \infty$ the tunneling time reduces as the number of barrier increases. This is also shown graphically in Fig \ref{tt_graph_n_barrier}-b. It is observed that (Fig \ref{tt_graph_n_barrier}-b) that the tunneling time first attain a maxima and then monotonically reduces. The monotonic reduction of the tunneling time with the thickness of the barrier after a certain thickness is a further paradoxial result and needs further investigation.

{\it \bf{Acknowledgements}}: \\
Nearly all the work of this manuscript was done when MH was in Space Astronomy Group (SAG) /URSC before moving to SSPO/ISRO HQ. MH acknowledges supports from SAG and Director-SSPO for the encouragement of research activities. BPM acknowledges the support from CAS, Department of Physics, BHU.

\end{document}